\documentclass[sigconf]{acmart}

\usepackage{booktabs} 
\usepackage{amsmath}
\usepackage{bm}
\usepackage[T1]{fontenc}

\usepackage{balance}
\usepackage{mathtools}
\DeclarePairedDelimiter{\ceil}{\lceil}{\rceil}
\usepackage{float}

\def\BibTeX{{\rm B\kern-.05em{\sc i\kern-.025em b}\kern-.08emT\kern-.1667em\lower.7ex\hbox{E}\kern-.125emX}}

\usepackage{subcaption}
\usepackage{float}

\copyrightyear{2020}
\acmYear{2020}
\setcopyright{acmlicensed}\acmConference[SIGIR '20]{Proceedings of the 43rd International ACM SIGIR Conference on Research and Development in Information Retrieval}{July 25--30, 2020}{Virtual Event, China}
\acmBooktitle{Proceedings of the 43rd International ACM SIGIR Conference on Research and Development in Information Retrieval (SIGIR '20), July 25--30, 2020, Virtual Event, China}
\acmPrice{15.00}
\acmDOI{10.1145/3397271.3401060}
\acmISBN{978-1-4503-8016-4/20/07}

\author{Casper Hansen}
\authornote{Both authors share the first authorship}
\affiliation{
  \city{University of Copenhagen}
}
\email{c.hansen@di.ku.dk}
\author{Christian Hansen}
\authornotemark[1]
\affiliation{
  \city{University of Copenhagen}
}
\email{chrh@di.ku.dk}
\author{Jakob Grue Simonsen}
\affiliation{
  \city{University of Copenhagen}
}
\email{simonsen@di.ku.dk}
\author{Stephen Alstrup}
\affiliation{
  \city{University of Copenhagen}
}
\email{s.alstrup@di.ku.dk}
\author{Christina Lioma}
\affiliation{
  \city{University of Copenhagen}
}
\email{c.lioma@di.ku.dk}
\begin{document}
\fancyhead{}
\begin{abstract}
Content-aware recommendation approaches are essential for providing meaningful recommendations for \textit{new} (i.e., \textit{cold-start}) items in a recommender system. We present a content-aware neural hashing-based collaborative filtering approach (NeuHash-CF), which generates binary hash codes for users and items, such that the highly efficient Hamming distance can be used for estimating user-item relevance. NeuHash-CF is modelled as an autoencoder architecture, consisting of two joint hashing components for generating user and item hash codes. Inspired from semantic hashing, the item hashing component generates a hash code directly from an item's content information (i.e., it generates cold-start and seen item hash codes in the same manner). This contrasts existing state-of-the-art models, which treat the two item cases separately. The user hash codes are generated directly based on user id, through learning a user embedding matrix. We show experimentally that NeuHash-CF significantly outperforms state-of-the-art baselines by up to 12\% NDCG and 13\% MRR in cold-start recommendation settings, and up to 4\% in both NDCG and MRR in standard settings where all items are present while training. Our approach uses 2-4x shorter hash codes, while obtaining the same or better performance compared to the state of the art, thus consequently also enabling a notable storage reduction.
\end{abstract}
\title{Content-aware Neural Hashing for Cold-start Recommendation}
\maketitle

\section{Introduction}
Personalizing recommendations is a key factor in successful recommender systems, and thus is of large industrial and academic interest. Challenges arise both with regards to efficiency and effectiveness, especially for large-scale systems with tens to hundreds of millions of items and users.

Recommendation approaches based on collaborative filtering (CF), content-based filtering, and their combinations have been investigated extensively (see the surveys in \cite{adomavicius2005toward,shi2014collaborative}), with CF based systems being one of the major methods in this area. CF based systems learn directly from either implicit (e.g., clicks) or explicit feedback (e.g., ratings), where matrix factorization approaches have traditionally worked well \cite{bennett2007netflix,Koren:2009:MFT:1608565.1608614}. CF learns $m$-dimensional user and item representations based on a factorization of the interaction-matrix between users and items, e.g., based on their click or rating history, such that the inner product can be used for computing user-item relevance. However, in the case of new unseen items (i.e., \emph{cold-start} items), standard CF methods are unable to learn meaningful representations, and thus cannot recommend those items (and similarly for cold-start users). To handle these cases, content-aware approaches are used when additional content information is available, such as textual descriptions, and have been shown to improve upon standard CF based methods \cite{lian2015content}. 

In large-scale recommendation settings, providing top-K recommendations among all existing items using an inner product is computationally costly, and thus provides a practical obstacle in employing these systems at scale. Hashing-based approaches solve this by generating \emph{binary} user and item hash codes, such that user-item relevance can be computed using the Hamming distance (i.e., the number of bit positions where two bit strings are different). The Hamming distance has a highly efficient hardware-level implementation, and has been shown to allow for real-time retrieval among a billion items \cite{shan2018recurrent}. Early work on hashing-based collaborative filtering systems \cite{karatzoglou2010collaborative,zhou2012learning,zhang2014preference} learned real-valued user and item representations, which were then in a later step discretized into binary hash codes. Further work focuses on end-to-end approaches, which improve upon the two-stage approaches by reducing the discretizing error by optimizing the hash codes directly \cite{zhang2016discrete,Liu:2019:CCC:3331184.3331206}. Recent content-aware hashing-based approaches \cite{Lian:2017:DCM:3097983.3098008,Zhang:2018:DDL:3159652.3159688} have been shown to perform well in both standard and cold-start settings, however they share the common problem of generating cold-start item hash codes differently from standard items, which we claim is unnecessary and limits their generalizability in cold-start settings.

We present a novel neural approach for content-aware hashing-based collaborative filtering (NeuHash-CF) robust to cold-start recommendation problems. NeuHash-CF consists of two joint hashing components for generating user and item hashing codes, which are connected in a variational autoencoder architecture. Inspired by semantic hashing \cite{salakhutdinov2009semantic}, the item hashing component learns to directly map an item's content information to a hash code, while maximizing its ability to reconstruct the original content information input. The user hash codes are generated directly based on the user's id through learning a user embedding matrix, and are jointly optimized with the item hash codes to optimize the log likelihood of observing each user-item rating in the training data. Through this end-to-end trainable architecture, all item hash codes are generated in the same way, independently of whether they are seen or not during training. We experimentally compare our NeuHash-CF to state-of-the-art baselines, where we obtain significant performance improvements in cold-start recommendation settings by up to 12\% NDCG and 13\% MRR, and up to 4\% in standard recommendation settings. Our NeuHash-CF approach uses 2-4x fewer bits, while obtaining the same or better performance than the state of the art, and notable storage reductions. 

In summary, we \textbf{contribute} a novel content-aware hashing-based collaborative filtering approach (NeuHash-CF), which in contrast to existing state-of-the-art approaches generates item hash codes in a unified way (not distinguishing between standard and cold-start items). 

\section{Related Work} \label{s:relwork}
The seminal work of \citet{das2007google} used a Locality-Sensitive Hashing \cite{gionis1999similarity} scheme, called Min-Hashing, for efficiently searching Google News, where a Jaccard measure for item-sharing between users was used to generate item and user hash codes. Following this, \citet{karatzoglou2010collaborative} used matrix factorization to learn real-valued latent user and item representations, which were then mapped to binary codes using random projections. Inspired by this, \citet{zhou2012learning} applied iterative quantization \cite{gong2012iterative} as a way of rotating and binarizing the real-valued latent representations, which had originally been proposed for efficient hashing-based image retrieval.
However, since the magnitude of the original real-valued representations are lost in the quantization, the Hamming distance between two hash codes might not correspond to the original relevance (inner product of real-valued vectors) of an item to a user. To solve this, \citet{zhang2014preference} imposed a constant norm constraint on the real-valued representations followed by a separate quantization.

Each of the above approaches led to improved recommendation performance, however, they can all be considered two-stage approaches, where the quantization is done as a post-processing step, rather than being part of the hash code learning procedure. Furthermore, post-processing quantization approaches have been shown to lead to large quantization errors \cite{zhang2016discrete}, leading to the investigation of approaches learning the hash codes directly.

Next, we review (1) hashing-based approaches for recommendation with explicit feedback; (2) content-aware hashing-based recommendation approaches designed for the cold-start setting of item recommendation; and (3) the related domain of semantic hashing, which our approach is partly inspired from. 

\subsection{Learning to Hash Directly}
Discrete Collaborative Filtering (DCF) \cite{zhang2016discrete} was the first approach towards learning item and user hash codes directly, rather than through a two-step approach. DCF is based on a matrix factorization formulation with additional constraints enforcing the discreteness of the generated hash codes. DCF further investigated balanced and de-correlation constraints to improve generalization by better utilizing the Hamming space. Inspired by DCF, \citet{zhang2017discrete} proposed Discrete Personalized Ranking (DPR) as a method designed for collaborative filtering with implicit feedback (in contrast to explicit feedback in the DCF case). DPR optimized a ranking objective through AUC and regularized the hash codes using both balance and de-correlation constraints similar to DCF. While these and previous two-stage approaches have led to highly efficient and improved recommendations, they are still inherently constrained by the limited representational ability of binary codes (in contrast to real-valued representations). To this end, Compositional Coding for Collaborative Filtering (CCCF) \cite{Liu:2019:CCC:3331184.3331206} was proposed as a hybrid approach between discrete and real-valued representations. CCCF considers each hash code as consisting of a number of blocks, each of which is associated with a learned real-valued scalar weight. The block weights are used for computing a \emph{weighted} Hamming distance, following the intuition that not all parts of an item hash code are equally relevant for all users. While this hybrid approach led to improved performance, it has a significant storage overhead (due to each hash code's block weights) and computational runtime increase, due to the weighted Hamming distance, compared to the efficient hardware-supported Hamming distance. 

\subsection{Content-aware Hashing}
A common problem for collaborative filtering approaches, both binary and real-valued, is the cold-start setting, where a number of items have not yet been seen by users.
In this setting, approaches based solely on traditional collaborative filtering cannot generate representations for the new items. Inspired by DCF, Discrete Content-aware Matrix Factorization (DCMF) \cite{Lian:2017:DCM:3097983.3098008} was the first hashing-based approach that also handled the cold-start setting. DCMF optimizes a multi-objective loss function, which most importantly learns hash codes directly for minimizing the squared rating error. Secondly, it also learns a latent representation for each content feature (e.g., each word in the content vocabulary), which is multiplied by the content features to approximate the learned hash codes, such that this can be used for generating hash codes in a cold-start setting. DCMF uses an alternating optimization strategy and, similarly to DCF, includes constraints enforcing bit balancing and de-correlation. Another approach, Discrete Deep Learning (DDL) \cite{Zhang:2018:DDL:3159652.3159688} learns hash codes similarly to DCMF, through an alternating optimization strategy solving a relaxed optimization problem. However, instead of learning latent representations for each content feature to solve the cold-start problem, they train a deep belief network \cite{hinton2006fast} to approximate the already learned hash codes based on the content features. This is a problem as described below.

DCMF and DDL both primarily learn hash codes not designed for cold-start settings, but then as a sub-objective learn how to map content features to new compatible hash codes for the cold-start setting. In practice, this is problematic as it corresponds to learning cold-start item hash codes based on previously learned hash codes from standard items, which we claim is unnecessary and limits their generalizability in cold-start settings.
In contrast, our proposed NeuHash-CF approach does not distinguish between between the settings for generating item hash codes, but rather always bases the item hash codes on the content features through a variational autoencoder architecture. As such, our approach can learn a better mapping from content features to hash code, since it is learned directly, as opposed to learning it in two steps by approximating the existing hash codes that have already been generated.

\subsection{Semantic Hashing}
The related area of Semantic Hashing \cite{salakhutdinov2009semantic} aims to map objects (e.g., images or text) to hash codes, such that similar objects have a short Hamming distance between them. Early work focused on two-step approaches based on learning real-valued latent representations followed by a rounding stage \cite{weiss2009spectral,zhang2010laplacian,zhang2010self}. Recent work has primarily used autoencoder-based approaches, either with a secondary rounding step \cite{chaidaroon2017variational}, or through direct optimization of binary codes using Bernoulli sampling and straight-through estimators for back-propagation during training \cite{shen2018nash,Hansen:2019:UNG:3331184.3331255,hansen-semhash-sigir-2020}. We draw inspiration from the latter approaches in the design of the item hashing component of our approach, as substantial performance gains have previously been observed in the semantic hashing literature over rounding-based approaches.

\section{Hashing-based Collaborative Filtering}
Collaborative Filtering learns real-valued latent user and item representations, such that the inner product between a user $u$ and item $i$ corresponds to the item's relevance to that specific user, where the ground truth is denoted as a user-item rating $R_{u,i}$.
Hashing-based collaborative filtering learns \emph{hash codes}, corresponding to \emph{binary} latent representations, for users and items. We denote $m$-bit user and item hash codes as $z_u \in \{-1, 1\}^m$ and $z_i \in \{-1, 1\}^m$, respectively. For estimating an item's relevance to a specific user in the hashing setting, the Hamming distance is computed as opposed to the inner product, as:
\begin{align}
    H(z_u, z_i) &= \sum_{j=1}^m {1}_{\big[ z_u^{(j)} \neq z_i^{(j)} \big]}
                =\text{SUM}\big( z_u \; \text{XOR} \; z_i \big)
\end{align}
Thus, the Hamming distance corresponds to summing the differing bits between the codes, which can be implemented very efficiently using hardware-level bit operations through the bitwise XOR and \emph{popcount} operations. The relation between the inner product and Hamming distance of hash codes is simply:
\begin{align}\label{eq:dotprod-ham}
    z_u^T z_i &= m - 2H(z_u, z_i)
\end{align}
meaning it is trivial to replace real-valued user and item representations with learned hash codes in an existing recommender system.

\subsection{Content-aware Neural Hashing-based Collaborative Filtering (NeuHash-CF)}
We first give an overview of our model, Content-aware Neural Hashing-based Collaborative Filtering (NeuHash-CF), and then detail its components. NeuHash-CF consists of two joint components for generating user and item hash codes. 
The item hashing component learns to derive item hash codes directly from the content features associated with each item. The item hashing component has two optimization objectives: (1) to maximize the likelihood of the observed user-item ratings, and (2) the unsupervised objective of reconstructing the original content features. Through this design, all item hash codes are based on content features, thus directly generating hash codes usable for both standard and cold-start recommendation settings. This contrasts existing state-of-the-art models \cite{Zhang:2018:DDL:3159652.3159688,Lian:2017:DCM:3097983.3098008} that separate how standard and cold-start item hash codes are generated. Through this choice, NeuHash-CF can generate higher quality cold-start item hash codes, but also improve the representational power of already observed items by better incorporating content features.

The user hashing component learns user hash codes, located within the same Hamming space as the item hash codes, by maximizing the likelihood of the observed user-item ratings, which is a shared objective with the item hashing component. Maximizing the likelihood of the observed user-item ratings influences the model optimization in relation to both user and item hash codes, while the unsupervised feature reconstruction loss of the item hashing component is focused only on the item hash codes. The aim of this objective combination is to ensure that the hash code distances enforce user-item relevance, but also that items with similar content have similar hash codes.

\begin{figure}
    \centering
    \includegraphics[width=0.85\linewidth]{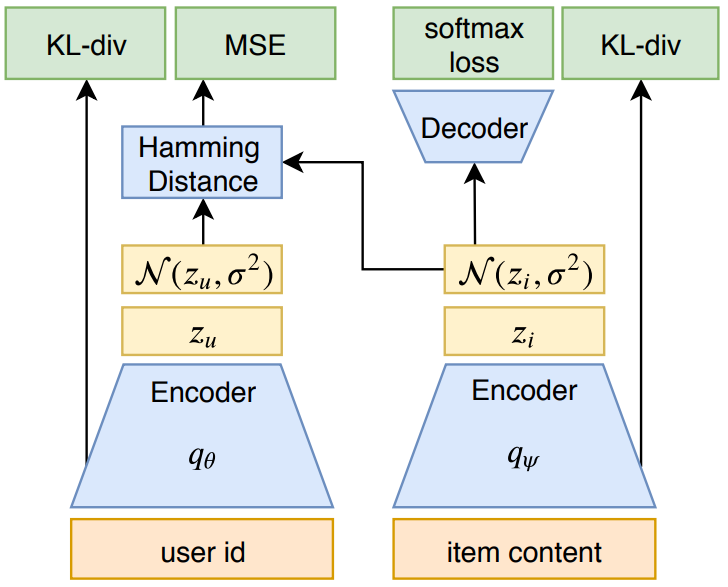}
    \vspace{-9pt}
    \caption{NeuHash-CF model overview.}
    \label{fig:neuhash}
    \vspace{-5pt}
\end{figure}

Next, we describe the architecture of our variational autoencoder (Section \ref{ss:framework}), followed by how users and items are encoded into hash codes (Section \ref{ss:encoder-functions}), decoded for obtaining a target value (Section \ref{ss:decoder-functions}), and lastly the formulation of the final loss function (Section \ref{ss:combined-loss}). We provide a visual overview of our model in Figure \ref{fig:neuhash}.

\subsection{Variational Autoencoder Architecture} \label{ss:framework}
We propose a variational autoencoder architecture for generating user and item hash codes, where we initially define the likelihood functions of each user and item as:
\begin{align}
    p(u) &= \prod_{i \in \mathbb{I}_u} p(R_{u,i}) \label{eq:pu} \\
    p(i) &= p(c_i) + \prod_{u \in \mathbb{U}_i} p(R_{u,i}) \label{eq:pi}
\end{align}
where $\mathbb{I}_u$ is the set of all items rated by user $u$, $\mathbb{U}_i$ is the set of all users who have rated item $i$, and $p(c_i)$ is the probability of observing the content of item $i$. We denote as $c_i \in \mathbb{R}$ the $n$-dimensional content feature vector (a bag-of-words representation) associated with each item, and denote the non-zero entries as $\mathbb{W}_{c_i}$. Thus, we can define the content likelihood similar to Eq. \ref{eq:pu} and \ref{eq:pi}:
\begin{align}
   p(c_i) = \prod_{w \in \mathbb{W}_{c_i}} p(w). \label{eq:pc}
\end{align}
In order to maximize the likelihood of the users and items, we need to maximize the likelihood of the observed ratings, $p(R_{u,i})$, as well as the word probabilities $p(w)$. Since they must be maximized based on the generated hash codes, we assume that $p(R_{u,i})$ is conditioned on both $z_u$ and $z_i$, and that $p(w)$ is conditioned on $z_i$. For ease of derivation, we choose to maximize the log likelihood instead of the raw likelihoods, such that the log likelihood of the observed ratings and item content can be computed as:
\begin{align}
    \log p(R_{u,i}) &= \log \sum_{z_i, z_u \in \{-1, 1\}^m} p(R_{u,i}|z_i, z_u) p(z_i) p(z_u) \\
    \log p(c_i) &=  \log \sum_{z_i \in \{-1, 1\}^m} p(c_i|z_i) p(z_i) 
\end{align}
where the hash codes are sampled by repeating $m$ consecutive Bernoulli trials, which as a prior is assumed to have equal probability of sampling either 1 or -1. Thus, $p(z_i)$ and $p(z_u)$ can be computed simply as:
\begin{align}
p(z) = \prod_{j=1}^m p^{\delta_j}(1 - p)^{1 - \delta_j}, \;\; \delta_j = 1_{\big[ z^{(j)} > 0 \big]}
\end{align}
where $z^{(j)}$ is the j'th bit of a hash code (either user or item), and where we set $p=0.5$ for equal sampling probability of 1 and -1. However, optimizing the log likelihoods directly is intractable, so instead we maximize their variational lower bounds \cite{kingma2014auto}:
\begin{align}
    \log p(R_{u,i}) &\geq E_{q_\psi, q_\theta} \big[ \log p(R_{u,i}|z_i, z_u) \big] \nonumber \\
    &- \text{KL}(q_\psi(z_i|i) || p(z_i)) - \text{KL}(q_\theta(z_u|u) || p(z_u)) \label{eq:elbo_rating} \\
    \log p(c_i) &\geq E_{q_\psi} \big[ \log p(c_i|z_i) \big] - \text{KL}(q_\psi(z_i|c_i) || p(z_i)) \label{eq:elbo_content}
\end{align}
where $q_\psi(z_i|i)$ and $q_\theta(z_u|u)$ are \emph{learned} approximate posterior probability distributions (see Section \ref{ss:encoder-functions}), and KL is the Kullback-Leibler divergence. Intuitively, the conditional log likelihood within the expectation term can be considered a reconstruction term, which represents how well either the observed ratings or item content can be decoded from the hash codes (see Section \ref{ss:decoder-functions}). The KL divergence can be considered as a regularization term, by punishing large deviations from the Bernoulli distribution with equal sampling probability of 1 and -1, which is computed analytically as:
\begin{align}
    \text{KL}(q_\psi(z_i|c_i) || p(z_i)) &=  q_\psi(c_i) \log \frac{q_\psi(c_i)}{p} \nonumber \\
    &+ (1-q_\psi(c_i)) \log \frac{1-q_\psi(c_i)}{p}
\end{align}
with $p=0.5$ for equal sampling probability. The KL divergence is computed similarly for the user hash codes using $\theta$. Next we describe how to compute the learned approximate posterior probability distributions.

\subsection{Encoder Functions} \label{ss:encoder-functions}
The learned approximate posterior distributions $q_\psi$ and $q_\theta$ can be considered encoder functions for items and users, respectively, and are both modeled through a neural network formulation. Their objective is to transform users and items into $m$ bit hash codes. 

\subsubsection{Item encoding}
An item $i$ is encoded based on its content $c_i$ through multiple layers to obtain sampling probabilities for generating the hash code:
\begin{align}
    l_1 &= \text{ReLU}\big( W_1 (c_i \odot w_{\text{imp}}) + b_1  \big) \label{eq:itemenc1} \\
    l_2 &= \text{ReLU}\big( W_2 l_1 + b_2  \big) \label{eq:itemenc2}
\end{align}
where $W$ and $b$ are learned weights and biases, $\odot$ is elementwise multiplication, and $w_{\text{imp}}$ is a learned importance weight for scaling the content words, which has been used similarly for semantic hashing \cite{Hansen:2019:UNG:3331184.3331255}. 
Next, we obtain the sampling probabilities by transforming the last layer, $l_2$, into an $m$-dimensional vector:
\begin{align}
    q_\psi(c_i) = \sigma\big( W_3 l_2 + b_3 \big)
\end{align}
where $\sigma$ is the sigmoid function to scale the output between 0 and 1, and $\psi$ is the set of parameters used for the item encoding. We can now sample the item hash code from a Bernoulli distribution, which can be computed for each bit as:
\begin{align}
    z_i^{(j)} = 2\ceil{ q_\psi(i)^{(j)} - \mu^{(j)} }-1
\end{align}
where $\mu \in [0,1]^m$ is an $m$-dimensional vector with uniformly sampled values. The model is trained using randomly sampled $\mu$ vectors, since it encourages model exploration because the same item may be represented as multiple different hash codes during training. However, to produce a deterministic output for testing once the model is trained, we fix each value within $\mu$ to 0.5 instead of a randomly sampled value.

\subsubsection{User encoding}
The user hash codes are learned similarly to the item hash codes, 
however, since we do not have a user feature vector, the hash codes are learned using only the user id. Thus, the sampling probabilities are learned as:
\begin{align}
    q_\theta(u) = \sigma \big( E_{\text{user}} 1_u \big)
\end{align}
where $E_{\text{user}} \in \mathbb{R}^{|U| \times m} $ is the learned user embedding, and $1_u$ is a one-hot encoding of user $u$. Following the same approach as the item encoding, we can sample the user hash code based on $q_\theta(u)$ for each bit as:
\begin{align}
    z_u^{(j)} = 2\ceil{ q_\theta(u)^{(j)} - \mu^{(j)} }-1
\end{align}
where $\theta$ is the set of parameters for user encoding. During training and testing, we use the same sampling strategy as for the item encoding. For both users and items, we use a straight-through estimator \cite{bengio2013estimating} for computing the gradients for backpropgation through the sampled hash codes.

\subsection{Decoder Functions} \label{ss:decoder-functions}
\subsubsection{User-item rating decoding}
The first decoding step aims to reconstruct the original user-item rating $R_{u,i}$, which corresponds to computing the conditional log likelihood of Eq. \ref{eq:elbo_rating}, i.e., $\log p(R_{u,i)}|z_i,z_u)$. We first transform the user-item rating into the same range as the inner product between the hash codes:
\begin{align}
    \hat{R}_{u,i} = 2 m \frac{R_{u,i}}{\text{max rating}} - m
\end{align}
Similarly to \cite{liang2018variational,sachdeva2019sequential}, we assume the ratings are Gaussian distributed around their true mean for each rating value, such that we can compute the conditional log likelihood as:
\begin{align}
   \log p(R_{u,i}|z_i,z_u) = \log \mathcal{N}\big( \hat{R}_{u,i} - z_i^T z_u, \sigma^2 \big) \label{eq:logp}
\end{align}
where the variance $\sigma^2$ is constant, thus providing an equal weighting of all ratings. However, the exact value of the variance is irrelevant, since maximizing Eq. \ref{eq:logp} corresponds to simply minimizing the squared error (MSE) of the mean term, i.e., $\hat{R}_{u,i} - z_i^T z_u$. Thus, maximizing the log likelihood is equivalent to minimizing the MSE, as similarly done in related work \cite{zhang2016discrete,Lian:2017:DCM:3097983.3098008,Zhang:2018:DDL:3159652.3159688}. Lastly, note that due to the equivalence between the inner product and the Hamming distance (see Eq. \ref{eq:dotprod-ham}), this directly optimizes the hash codes for the Hamming distance. 

\subsubsection{Item content decoding}
The secondary decoding step aims to reconstruct the original content features given the generated item hash code in Eq. \ref{eq:elbo_content}, i.e., $\log p(c_i|z_i)$. We compute this as the summation of word log likelihoods (based on Eq. \ref{eq:pc}) using a softmax:
\begin{align}
    \log p(c_i|z_i) = \sum_{ w \in \mathbb{W}_{c_i} } \log \frac{e^{z_i^T (E_{\text{word}}( 1_w \odot w_{\text{imp}})) + b_w}}{e^{ \sum_{w' \in \mathbb{W}} z_i^T (E_{\text{word}}( 1_{w'} \odot w_{\text{imp}})) + b_{w'} }}
\end{align}
where $1_w$ is a one-hot encoding for word $w$, $\mathbb{W}$ is the set of all vocabulary words of the content feature vectors, $E_{\text{word}} \in \mathbb{R}^{|\mathbb{W}| \times m}$ is a learned word embedding, $b_w$ is a word-level bias term, and the learned importance weight $w_{\text{imp}}$ is the same as in Eq. \ref{eq:itemenc1}. This softmax expression is maximized when the item hash codes are able to decode the original content words.

\subsubsection{Noise infusion for robustness} 
Previous work on semantic hashing has shown that infusing random noise into the hash codes before decoding increases robustness, and leads to more generalizable hash codes \cite{shen2018nash,chaidaroon2018deep,Hansen:2019:UNG:3331184.3331255}. Thus, we apply a Gaussian noise to both user and item hash codes before decoding:
\begin{align}
    \text{noise}(z, \sigma^2) = z + \epsilon \sigma^2, \;\; \epsilon \sim \mathcal{N}(0, I)
\end{align}
where variance annealing is used for decreasing the initial value of $\sigma^2$ in each training iteration.

\subsection{Combined Loss Function} \label{ss:combined-loss}
NeuHash-CF can be trained in an end-to-end fashion by maximising the combination of the variational lower bounds from Eq. \ref{eq:elbo_rating} and \ref{eq:elbo_content}, corresponding to the following loss:
\begin{align} \label{eq:final-loss}
    \mathcal{L} = \mathcal{L}_{\text{rating}} + \alpha \mathcal{L}_{\text{content}}
\end{align}
where $\mathcal{L}_{\text{rating}}$ corresponds to the lower bound in Eq. \ref{eq:elbo_rating}, $\mathcal{L}_{\text{content}}$ corresponds to the lower bound in Eq. \ref{eq:elbo_content}, and $\alpha$ is a tunable hyper parameter to control the importance of decoding the item content. 

\section{Experimental Evaluation}
\subsection{Datasets}
We evaluate our approach on well-known and publicly available datasets with explicit feedback, where we follow the same preprocessing as related work \cite{Lian:2017:DCM:3097983.3098008,Zhang:2018:DDL:3159652.3159688,wang2011collaborative} as described in the following. We disallow users to have rated the same item multiple times and use only the last rating in these cases. Due to the very high sparsity of these types of datasets, we apply a filtering to densify the dataset. We remove users who have rated fewer then 20 items, as well items that have been rated by fewer than 20 users. Since the removal of either a user or item may violate the density requirements, we apply the filtering iteratively until all users and items satisfy the requirement.
The datasets are described below and summarized in Table \ref{tab:dataset-stats}:
\begin{description}
    \item[\textbf{Yelp}] is from the Yelp Challenge\footnote{\url{https://www.yelp.com/dataset/challenge}}, which consists of user ratings and textual reviews on locations such as hotels, restaurants, and shopping centers. User ratings range between 1 (worst) to 5 (best), and most ratings are associated with a textual review.
    
    \item[\textbf{Amazon} \cite{he2016ups}] is from a collection of book reviews from Amazon\footnote{\url{http://jmcauley.ucsd.edu/data/amazon/}}. Similarly to Yelp, each user rates a number of books between 1 to 5, and most are accompanied by a textual review as well.
\end{description}
Similarly to related work \cite{Lian:2017:DCM:3097983.3098008,Zhang:2018:DDL:3159652.3159688,wang2011collaborative}, to obtain content information related to each item, we use the textual reviews (when available) by users for an item. We filter stop words and aggregate all textual reviews for each item into a single large text, and compute the TF-IDF bag-of-words representations, where the top 8000 unique words are kept as the content vocabulary. We apply this preprocessing step separately on each dataset, thus resulting in two different vocabularies.

\begin{table}[]
    \centering
        \caption{Dataset statistics after preprocessing such that each user has at least rated 20 items, and each item has at least been rated by 20 users.}
    \begin{tabular}{lccccc}
    \toprule
        Dataset & \#users & \#items & \#ratings & sparsity \\ \midrule
        Yelp & 27,147 & 20,266 & 1,293,247 & 99.765\% \\
        Amazon & 35,736 & 38,121 & 1,960,674 & 99.856\% \\
        \bottomrule
    \end{tabular}
    \label{tab:dataset-stats}
    \vspace{-10pt}
\end{table}

\subsection{Experimental Design} \label{ss:eval-framework}
Following \citet{wang2011collaborative}, we use two types of recommendations settings: 1) in-matrix regression for estimating the relevance of known items with existing ratings, and 2) out-of-matrix regression for estimating the relevance of cold-start items. Both of these recommendation types lead to different evaluation setups as described next.

\subsubsection{In-matrix regression}
In-matrix regression can be considered the standard setup of all items (and users) being known at all times, and thus corresponds to the setting solvable by standard collaborative filtering. We split each user's items into a training and testing set using a 50/50 split, and use 15\% of the training set as a validation set for hyper parameter tuning.

\subsubsection{Out-of-matrix regression} \label{sss:outreg}
Out-of-matrix regression is also known as a cold-start setting, where new items are to be recommended. In comparison to in-matrix regression, this task cannot be solved by standard collaborative filtering. We sort all items by their number of ratings, and then proportionally split them 50/50 into a training and testing set, such that each set has approximately the same number of items with similar number of ratings. Similarly to the in-matrix regression setting, we use 15\% of the training items as a validation set for hyper parameter tuning. 

\subsection{Evaluation Metrics}
We evaluate the effectiveness of our approach and the baselines as a ranking task with the aim of placing the most relevant (i.e., highest rated) items at the top of a ranked list. As detailed in Section \ref{ss:eval-framework}, each user has a number of rated items, such that the ranked list is produced by sorting each user's testing items by their Hamming distance between the user and item hash codes. To measure the quality of the ranked list, we use Normalized Discounted Cumulative Gain (NDCG), which incorporates both ranking precision and the position of ratings. 
Secondly, we are interested in the first position of the item with the highest rating, as this ideally should be in the top. To this end, we compute the Mean Reciprocal Rank (MRR) of the highest ranked item with the highest given rating from the user's list of testing items.

\subsection{Baselines}
We compare NeuHash-CF against existing state-of-the-art content-aware hashing-based recommendation approaches, as well as hashing-based approaches that are not content-aware to highlight the benefit of including content:
\begin{description}
    \item[DCMF] Discrete Content-aware Matrix Factorization \cite{Lian:2017:DCM:3097983.3098008}\footnote{\url{https://github.com/DefuLian/recsys/tree/master/alg/discrete/dcmf}} is a content-aware matrix factorization technique, which is discretized and optimized through solving multiple mixed-integer subproblems. Similarly to our approach, its primary objective is to minimize the squared error between the rating and estimated rating based on the Hamming distance. It also learns a latent representation for each word in the text associated to each item, which is used for generating hash codes for cold-start items. 
    \item[DDL] Discrete Deep Learning \cite{Zhang:2018:DDL:3159652.3159688}\footnote{\url{https://github.com/yixianqianzy/ddl}} also uses an alternating optimizing strategy for solving multiple mixed-integer subproblems, where the primary objective is a mean squared error loss. In contrast to DCMF, DDL uses a deep belief network for generating cold-start item hash codes, which is trained by learning to map the content of known items into their hash codes generated in the first part of the approach. 
    \item[DCF] Discrete Collaborative Filtering \cite{zhang2016discrete}\footnote{\url{https://github.com/hanwangzhang/Discrete-Collaborative-Filtering}} can be considered the predecessor to DCMF, but is not content-aware, which was the primary novelty of DCMF. 
    \item[NeuHash-CF/no.C] We include a version of our NeuHash-CF that is not content-aware, which is done by simply learning item hash codes similarly to user hash codes, thus not including any content features.
\end{description}
For both DCMF and DDL, hash codes for cold-start items are seen as a secondary objective, as they are generated differently from non-cold-start item hash codes. In contrast, our NeuHash-CF treats all items identically as all item hash codes are generated based on content features alone. 

To provide a comparison to non-hashing based approaches, which are notably more computationally expensive for making recommendations (see Section \ref{ss:efficiency-storage}), we also include the following baselines:
\begin{description}
    \item[FM] Factorization Machines \cite{rendle2010factorization} works on a concatenated $n$-dimensional vector of the one-hot encoded user id, one-hot encoder item id, and the content features. It learns latent vectors, as well as scalar weights and biases for each of the $n$ dimensions. FM estimates the user-item relevance by computing a weighted sum of all non-zero entries and all interactions between non-zero entries of the concatenated vector. This results in a large amount of inner product computations and a large storage cost associated with the latent representations and scalars. We use the FastFM implementation \cite{JMLR:v17:15-355}\footnote{\url{https://github.com/ibayer/fastFM}}.
    \item[MF] Matrix Factorization \cite{Koren:2009:MFT:1608565.1608614} is a classic non-content-aware collaborative filtering approach, which learns real-valued item and user latent vectors, such that the inner product corresponds to the user-item relevance. MF is similar to a special case of FM without any feature interactions.
\end{description}

\begin{table*}[]
    \centering
    \caption{NDCG@k scores on in-matrix and out-of-matrix settings for the Amazon and Yelp datasets. Bold numbers represent the best hashing-based approach and statistically significant results compared to the best hashing-based baseline per column are marked with a star. Dashed lines correspond to not content-aware approaches in out-of-matrix setting.}
    \vspace{-6pt}
    \scalebox{0.825}{
    \begin{tabular}{@{}llll|lll|lll|lll|lll|lll@{}}
    \toprule
      & \multicolumn{9}{c|}{Yelp (in-matrix)} & \multicolumn{9}{c}{Yelp (out-of-matrix)}\\ 
      & \multicolumn{3}{c|}{16 dim.} & \multicolumn{3}{c|}{32 dim.} & \multicolumn{3}{c|}{64 dim.} &  \multicolumn{3}{c|}{16 dim.} & \multicolumn{3}{c|}{32 dim.} & \multicolumn{3}{c}{64 dim.} \\ 
      NDCG & @2 & @6 & @10 & @2 & @6 & @10 & @2 & @6 & @10 & @2 & @6 & @10 & @2 & @6 & @10 & @2 & @6 & @10 \\
      \toprule
NeuHash-CF & \textbf{.662}$^*$ & \textbf{.701}$^*$ & \textbf{.752}$^*$ & \textbf{.681}$^*$ & \textbf{.718}$^*$ & \textbf{.766}$^*$ & \textbf{.697}$^*$ & \textbf{.731}$^*$ & \textbf{.776}$^*$ &  \textbf{.646}$^*$ & \textbf{.694}$^*$ & \textbf{.747}$^*$ & \textbf{.687}$^*$ & \textbf{.725}$^*$ & \textbf{.772}$^*$ & \textbf{.702}$^*$ & \textbf{.737}$^*$ & \textbf{.780}$^*$ \\
DCMF & .642 & .678 & .733 & .655 & .691 & .743 & .670 & .701 & .752 &  .611 & .647 & .703 & .617 & .655 & .709 & .626 & .664 & .717 \\
DDL & .636 & .674 & .729 & .651 & .686 & .739 & .664 & .698 & .749 &  .575 & .615 & .673 & .579 & .622 & .681 & .612 & .646 & .700 \\
\midrule
NeuHash-CF/no.C & .634 & .672 & .727 & .655 & .689 & .741 & .666 & .699 & .749 &  - & - & - & - & - & - & - & - & - \\
DCF & .639 & .676 & .730 & .649 & .685 & .738 & .671 & .700 & .750 &  - & - & - & - & - & - & - & - & - \\
\midrule
MF (real-valued) & .755$^*$ & .763$^*$ & .800$^*$ & .755$^*$ & .763$^*$ & .800$^*$ & .755$^*$ & .763$^*$ & .800$^*$ &  - & - & - & - & - & - & - & - & - \\
FM (real-valued) & .754$^*$ & .763$^*$ & .801$^*$ & .750$^*$ & .760$^*$ & .798$^*$ & .744$^*$ & .755$^*$ & .794$^*$ &  .731$^*$ & .750$^*$ & .789$^*$ & .724$^*$ & .744$^*$ & .785$^*$ & .719$^*$ & .740$^*$ & .781$^*$ \\
\bottomrule
    \end{tabular}}
    \centering
    \scalebox{0.825}{
    \begin{tabular}{@{}llll|lll|lll|lll|lll|lll@{}}
      & \multicolumn{9}{c|}{Amazon (in-matrix)} & \multicolumn{9}{c}{Amazon (out-of-matrix)}\\ 
      & \multicolumn{3}{c|}{16 dim.} & \multicolumn{3}{c|}{32 dim.} & \multicolumn{3}{c|}{64 dim.} &  \multicolumn{3}{c|}{16 dim.} & \multicolumn{3}{c|}{32 dim.} & \multicolumn{3}{c}{64 dim.} \\ 
    NDCG & @2 & @6 & @10 & @2 & @6 & @10 & @2 & @6 & @10 & @2 & @6 & @10 & @2 & @6 & @10 & @2 & @6 & @10 \\
      \toprule
NeuHash-CF & \textbf{.759}$^*$ & \textbf{.777}$^*$ & \textbf{.810}$^*$ & \textbf{.780}$^*$ & \textbf{.798}$^*$ & \textbf{.827}$^*$ & \textbf{.786}$^*$ & \textbf{.803}$^*$ & \textbf{.831}$^*$ &  \textbf{.758}$^*$ & \textbf{.778}$^*$ & \textbf{.809}$^*$ & \textbf{.769}$^*$ & \textbf{.788}$^*$ & \textbf{.818}$^*$ & \textbf{.787}$^*$ & \textbf{.804}$^*$ & \textbf{.831}$^*$ \\
DCMF & .749 & .767 & .800 & .761 & .777 & .810 & .773 & .788 & .818 &  .727 & .748 & .782 & .729 & .749 & .784 & .733 & .752 & .786 \\
DDL & .734 & .755 & .791 & .748 & .768 & .802 & .762 & .779 & .811 &  .704 & .728 & .766 & .705 & .729 & .767 & .705 & .727 & .766 \\
\midrule
NeuHash-CF/no.C & .748 & .768 & .802 & .760 & .776 & .808 & .771 & .785 & .816 &  - & - & - & - & - & - & - & - & - \\
DCF & .745 & .767 & .802 & .759 & .776 & .809 & .774 & .787 & .818 &  - & - & - & - & - & - & - & - & - \\
\midrule
MF (real-valued) & .824$^*$ & .826$^*$ & .848$^*$ & .824$^*$ & .826$^*$ & .848$^*$ & .824$^*$ & .826$^*$ & .848$^*$ &  - & - & - & - & - & - & - & - & - \\
FM (real-valued) & .821$^*$ & .822$^*$ & .845$^*$ & .817$^*$ & .819$^*$ & .843$^*$ & .813$^*$ & .816$^*$ & .841$^*$ &  .792$^*$ & .800$^*$ & .827$^*$ & .785$^*$ & .793$^*$ & .821$^*$ & .780$^*$ & .790$^*$ & .819$^*$ \\
\bottomrule
    \end{tabular}}
    \label{tab:mainres_ndcg}
    \vspace{-6pt}
\end{table*}

\begin{table*}[]
    \centering
        \caption{MRR scores in both in-matrix and out-of-matrix settings. Bold numbers represent the best hashing-based approach and statistically significant results compared to the best hashing-based baseline per column are marked with a star. Dashed lines correspond to not content-aware approaches in out-of-matrix setting.}
        \vspace{-6pt}
    \scalebox{0.99}{
    \begin{tabular}{@{}llll|lll|lll|lll@{}}
    \toprule
      & \multicolumn{3}{c|}{Yelp (in-matrix)} & \multicolumn{3}{c|}{Yelp (out-of-matrix)} & \multicolumn{3}{c|}{Amazon (in-matrix)} & \multicolumn{3}{c}{Amazon (out-of-matrix)}\\ 
    MRR & \small 16 dim. & \small 32 dim. & \small 64 dim. & \small 16 dim. & \small 32 dim. & \small 64 dim. & \small 16 dim. & \small 32 dim. & \small 64 dim. & \small 16 dim. & \small 32 dim. & \small 64 dim. \\
      \toprule
NeuHash-CF & \textbf{.646}$^*$ & \textbf{.668}$^*$ & \textbf{.687}$^*$ & \textbf{.628}$^*$ & \textbf{.674}$^*$ & \textbf{.692}$^*$ & \textbf{.749}$^*$ & \textbf{.770}$^*$ & \textbf{.779}$^*$ & \textbf{.750}$^*$ & \textbf{.764}$^*$ & \textbf{.782}$^*$ \\
DCMF & .629 & .644 & .660 & .598 & .604 & .612 & .738 & .753 & .767 & .719 & .721 & .726 \\
DDL & .620 & .638 & .651 & .557 & .562 & .604 & .721 & .741 & .753 & .696 & .694 & .694 \\
\midrule
NeuHash-CF/no.C & .621 & .642 & .656 & - & - & - & .737 & .752 & .764 & - & - & - \\
DCF & .626 & .636 & .664 & - & - & - & .736 & .751 & .769 & - & - & - \\
\midrule
MF (real-valued) & .767$^*$ & .767$^*$ & .767$^*$ & - & - & - & .826$^*$ & .826$^*$ & .826$^*$ & - & - & - \\
FM (real-valued) & .761$^*$ & .756$^*$ & .750$^*$ & .730$^*$ & .722$^*$ & .717$^*$ & .824$^*$ & .821$^*$ & .815$^*$ & .792$^*$ & .784$^*$ & .780$^*$ \\ \bottomrule
    \end{tabular}}
    \label{tab:mainres_mrr}
    \vspace{-6pt}
\end{table*}
\subsection{Tuning}
For training our NeuHash-CF approach, we use the Adam \cite{kingma2014adam} optimizer with learning rates selected from $\{0.0005, 0.0001\}$ and batch sizes from $\{500, 1000, 2000\}$, where 0.0005 and 2000 were consistently chosen. We also tune the number of encoder layers from $\{1,2,3\}$ and the number of neurons in each from $\{500, 1000, 2000\}$; most runs had the optimal validation performance with 2 layers and 1000 neurons. To improve robustness of the codes we added Gaussian noise before decoding the hash codes, where the variance was initially set to 1 and decreased by 0.01\% after every batch. Lastly, we tune $\alpha$ in Eq. \ref{eq:final-loss} from $\{0.001, 0.01, 0.1\}$, where 0.001 was consistently chosen.
The code\footnote{We make the code publicly available at \url{https://github.com/casperhansen/NeuHash-CF}} is written in TensorFlow  \cite{abadi2016tensorflow}. For all baselines, we tune the hyper parameters on the validation set as described in the original papers.

\subsection{Results} \label{ss:results}
The experimental comparison is summarized in Table \ref{tab:mainres_ndcg} and \ref{tab:mainres_mrr} for NDCG@$\{2,6,10\}$ and MRR, respectively. The tables are split into in-matrix and out-of-matrix evaluation settings for both datasets, and the methods can be categorized into groups: (1) content-aware (NeuHash-CF, DCMF, DDL), (2) not content-aware (NeuHash-CF/no.C, DCF), (3) real-valued not content-aware (MF), and (4) real-valued content-aware (FM). For all methods, we compute hash codes (or latent representations for MF and FM) of length $m \in \{16,32,64\}$.
We use a two-tailed paired t-test for statistical significance testing against the best performing hashing-based baseline. Statistically significant improvements, at the 0.05 level, over the best performing hashing-based baseline per column are marked with a star ($^*$), and the best performing hashing-based approach is shown in bold.  

\subsubsection{In-matrix regression} 
In the in-matrix setting, where all items have been rated in the training data, our NeuHash-CF significantly outperforms all hashing-based baselines. On Yelp, we observe improvements in NDCG by up to 0.03, corresponding to a 4.3\% improvement. On Amazon, we observe improvements in NDCG by up to 0.02, corresponding to a 2.7\% improvement. Similar improvements are noted on both datasets for MRR (1.6-4.1\% improvements). On all datasets and across the evaluated dimensions, NeuHash-CF performs similarly or better than state-of-the-art hashing-based approaches while using 2-4 times fewer bits, thus providing both a significant performance increase as well as a 2-4 times storage reduction.
Interestingly, the performance gap between existing content-aware and not content-aware approaches is relatively small. When considering the relative performance increase of our NeuHash-CF with and without content features, we see the benefit of basing the item hash codes directly on the content. DCMF and DDL both utilize the content features for handling cold-start items, but not to the same degree for the in-matrix items, which we argue explains the primary performance increase observed for NeuHash-CF, since NeuHash-CF/no.C performs similarly to the baselines.

We also include MF and FM as real-valued baselines to better gauge the discretization gap. As expected, the real-valued approaches outperform the hashing-based approaches, however as the number of bit increases the performance difference decreases. This is to be expected, since real-valued approaches reach faster a potential representational limit, where more dimensions would not positively impact the ranking performance. In fact, for FM we observe a marginal performance drop when increasing its number of latent dimensions, thus indicating that it is overfitting. In contrast, MF keeps the same performance (differing on far out decimals) independently of its number of latent dimensions. 

\begin{table*}[]
    \centering
        \caption{NDCG@10 and MRR scores for 32 dimensional representations in varying cold-start scenarios with 10-50\% of the items used for training. Bold numbers represent the best hashing-based approach and statistically significant results compared to the best hashing-based baseline in each column are marked with a star.}
        \vspace{-9pt}
    \scalebox{0.76}{
    \begin{tabular}{@{}lll|ll|ll|ll|ll|ll|ll|ll|ll|ll@{}}
    \toprule
      &  \multicolumn{10}{c|}{Yelp (out-of-matrix)} & \multicolumn{10}{c}{Amazon (out-of-matrix)} \\ 
      & \multicolumn{2}{c}{10\%} & \multicolumn{2}{c}{20\%} & \multicolumn{2}{c}{30\%} & \multicolumn{2}{c}{40\%} & \multicolumn{2}{c|}{50\%} & \multicolumn{2}{c}{10\%} & \multicolumn{2}{c}{20\%} & \multicolumn{2}{c}{30\%} & \multicolumn{2}{c}{40\%} & \multicolumn{2}{c}{50\%} \\
    NDCG & \small @10 & \small MRR & \small @10 & \small MRR & \small @10 & \small MRR & \small @10 & \small MRR & \small @10 & \small MRR & \small @10 & \small MRR & \small @10 & \small MRR & \small @10 & \small MRR & \small @10 & \small MRR & \small @10 & \small MRR \\
    \toprule
NeuHash-CF & \textbf{.730}$^*$ & \textbf{.603}$^*$ & \textbf{.750}$^*$ & \textbf{.634}$^*$ & \textbf{.769}$^*$ & \textbf{.666}$^*$ & \textbf{.771}$^*$ & \textbf{.668}$^*$ & \textbf{.772}$^*$ & \textbf{.674}$^*$ & \textbf{.794}$^*$ & \textbf{.727}$^*$ & \textbf{.812}$^*$ & \textbf{.753}$^*$ & \textbf{.817}$^*$ & \textbf{.761}$^*$ & \textbf{.818}$^*$ & \textbf{.763}$^*$ & \textbf{.818}$^*$ & \textbf{.764}$^*$ \\
DCMF & .688 & .572 & .693 & .578 & .704 & .593 & .710 & .602 & .709 & .604 & .774 & .710 & .778 & .712 & .781 & .717 & .784 & .720 & .784 & .721 \\
DDL & .678 & .556 & .681 & .562 & .687 & .572 & .684 & .571 & .681 & .562 & .770 & .713 & .766 & .689 & .767 & .700 & .765 & .693 & .767 & .694 \\
\midrule
FM (real-valued) & .766$^*$ & .688$^*$ & .776$^*$ & .707$^*$ & .778$^*$ & .712$^*$ & .786$^*$ & .724$^*$ & .785$^*$ & .722$^*$ & .806$^*$ & .759$^*$ & .813$^*$ & .771$^*$ & .817$^*$ & .775$^*$ & .823$^*$ & .786$^*$ & .821$^*$ & .784$^*$ \\
\bottomrule
    \end{tabular}}

    \label{tab:cold_res}
    \vspace{-7pt}
\end{table*}

\subsubsection{Out-of-matrix regression} 
We now consider the out-of-matrix setting, corresponding to recommending cold-start items. NeuHash-CF significantly outperforms the existing state-of-the-art hashing-based baselines even more than for the in-matrix setting. On Yelp, we observe the smallest NDCG increase for 16 bit at 0.035, which is however doubled in most cases for 32 and 64 bits, corresponding to improvements of up to 12.1\% gain over state-of-the-art baselines. We observe a similar trend on Amazon, where the lowest improvement of 0.027 NDCG is observed at 16 bits, but increasing the number of bits leads to consistently larger improvements of up to 7.4\%. These results are also consistent with MRR, where increasing the number of bits provides increasingly larger performance increases between +5 and +13.1\% on Yelp and between +4.3 and +7.7\% on Amazon. In all cases, the performance of NeuHash-CF on 16 bits is even better than the best baseline at 64 bits, thus verifying the high quality of the hash codes generated by NeuHash-CF.

For the real-valued FM baseline, we observe that it outperforms ours and existing baselines at 16 and 32 dimensions, however at 64 dimensions NeuHash-CF outperforms FM on Amazon for NDCG@$\{6,10\}$ (across all dimensions). When we consider Yelp, NeuHash-CF obtains a NDCG@10 within 0.01 of FM, but worse on the other NDCG cut offs and on MRR. 

\subsubsection{Out-of-matrix regression with limited training data} 
To evaluate how the content-aware approaches generalize to the cold-start setting depending on the number of training items, we furthermore create smaller versions of the 50/50 out-of-matrix split used previously. In addition to using 50\% of the data for the training set, we consider splits using 10\%, 20\%, 30\%, and 40\% as well. In all out-of-matrix settings the validation and testing sets are identical to be able to compare the impact of the training size.
The results can be seen in Table \ref{tab:cold_res} for 32 bit hash codes and 32 latent dimensions in FM. Similarly to before, NeuHash-CF outperforms the hashing-based baselines in all cases with similar gains as observed previously. Most approaches, except DDL on Amazon, obtain the lowest performance using 10\% of the data, and more training items generally improve the performance, although at 30-50\% the pace of improvement slows down significantly. This indicates that the methods have observed close to sufficiently many training items and increasing the amount may not lead to better generalizability of the cold-start hash codes. Interestingly, NeuHash-CF obtains the largest improvement going from 10\% to 50\% on both NDCG and MRR, indicating that it generalizes better than the baselines. In contrast, DDL does not improve on Amazon by including more training items, which indicates that its ability to generalize to cold-start items is rather limited.

\begin{figure}
    \centering
    \scalebox{0.6}{
    \includegraphics[width=0.7\textwidth]{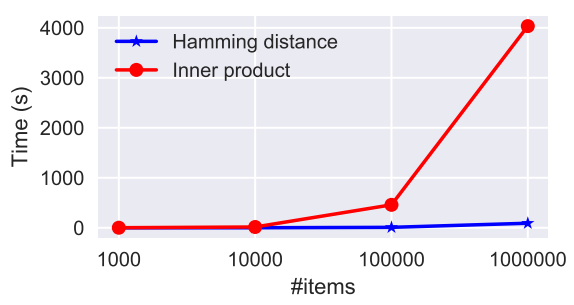}}
    \vspace{-13pt}
    \caption{Computation time for all Hamming distances and inner products for 100,000 users and up to 1,000,000 items.}
    \label{fig:eff}
\end{figure}
\subsection{Computational Efficiency} \label{ss:efficiency-storage}
To study the high efficiency of using hash codes in comparison to real-valued vectors, we consider a setup of 100,000 users and 1,000-1,000,000 items. We randomly generate hash codes and real-valued vectors and measure the time taken to compute all Hamming distances (or inner products) from each user to all items, resulting in a total $10^8$-$10^{11}$ computations. We use a machine with a 64 bit instruction set\footnote{We used an Intel Xeon CPU E5-2670}, and hence generate hash codes and vectors of length 64. We report the average runtime over 10 repetitions in Figure \ref{fig:eff}, and observe a speed up of a factor 40-50 for the Hamming distance, highlighting the efficiency benefit of hashing-based approaches. For FM, its dominating cost is its large number of inner product computations, which scales quadratically in the number of non-zero content features for a given item, thus making it highly intractable in large-scale settings.

\begin{figure*}[h]
    \centering
    \includegraphics[width=0.99\linewidth]{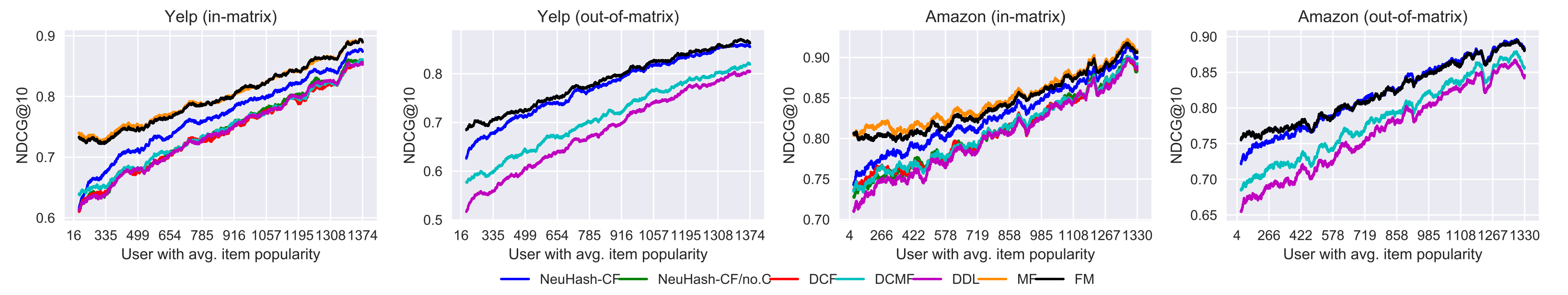}
    \vspace{-7pt}
    \caption{Impact of the average item popularity per user on NDCG@10 for 32 bit hash codes.}
    \vspace{-7pt}
    \label{fig:pop}
\end{figure*}
\begin{figure*}[h]
    \centering
    \includegraphics[width=0.99\linewidth]{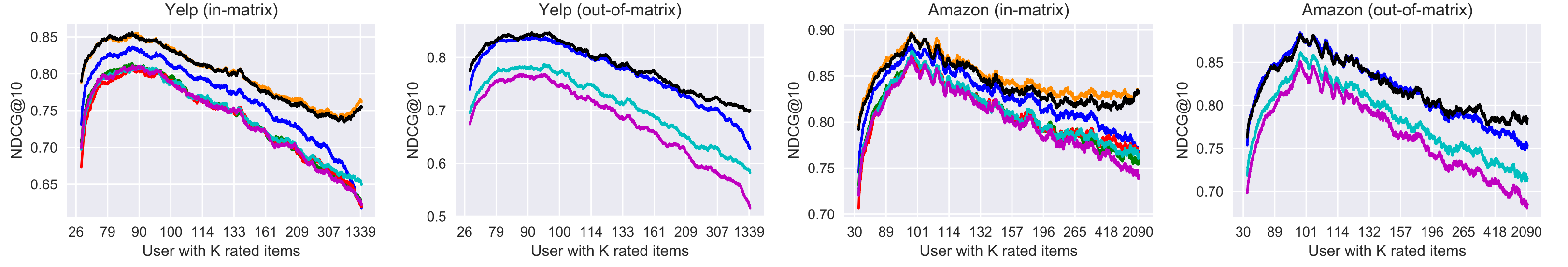}
    \vspace{-7pt}
    \caption{Impact of the number of items per user on NDCG@10 for 32 bit hash codes.}
    \label{fig:itemnum}
\end{figure*}

\subsection{Impact of Average Item Popularity per User} \label{ss:impact-avg-pop}
We now look at how different user characteristics impact the performance of the methods. We first compute the average item popularity of each user's list of rated items, and then order the users in ascending order of that average. An item's popularity is computed as the number of users who have rated that specific item, and thus the average item popularity of a user is representative of their attraction to popular content. Figure \ref{fig:pop} plots the NDCG@10 for 32 dimensional representations using a mean-smoothing window size of 1000 (i.e., each shown value is averaged based on the values within a window of 1000 users). Generally, all methods perform better for users who have a high average item popularity, where for Yelp we see a NDCG@10 difference of up to 0.25 from the lowest to highest average popularity (0.2 for Amazon). This observation can be explained by highly popular items occurring more times in the training data, such that they have a better learned representation. Additionally, the hashing-based approaches have a larger performance difference, compared to the real-valued MF and FM, which is especially due to their lower relative performance for users with a very low average item popularity (left side of plots). In the out-of-matrix setting the same trend is observed, however with our NeuHash-CF performing highly similarly to FM when excluding the users with the lowest average item popularity. We hypothesize that users with a low average item popularity have a more specialized preference, thus benefitting more from the higher representational power of real-valued representations.

\subsection{Impact of Number of Items per User}
We now consider how the number of items each user has rated impacts performance. We order users by their number of rated items and plot NDCG@10 for 32 bit hash codes. Figure \ref{fig:itemnum} plots this in the same way as in Figure \ref{fig:pop}. Generally across all methods, we observe that performance initially increases, but then drops once a user has rated close to 100 items, depending on the dataset. While the hashing-based approaches keep steadily dropping in performance, MF and FM do so at a slower pace and even increase for users with the highest number of rated items in the in-matrix setting. The plots clearly show that the largest performance difference, between the real-valued and hashing-based approaches, is for the group of users with a high number of rated items, corresponding to users with potentially the highest diversity of interests. In this setting, the limited representational power of hash codes, as opposed to real-valued representations, may not be sufficient to encode users with largely varied interests. We observe very similar trends for the out-of-matrix setting for cold-start items, although the performance gap between our NeuHash-CF and the real-valued approaches is almost entirely located among the users with a high number of rated items.


\section{Conclusion}
We presented content-aware neural hashing for collaborative filtering (NeuHash-CF), a novel hashing-based recommendation approach, which is robust to cold-start recommendation problems (i.e., the setting where the items to be recommended have not been rated previously). NeuHash-CF is a neural approach that consists of two joint components for generating user and item hash codes. The user hash codes are learned from an embedding based procedure using only the user's id, whereas the item hash codes are learned directly from associated content features (e.g., a textual item description). This contrasts existing state-of-the-art content-aware hashing-based methods \cite{Lian:2017:DCM:3097983.3098008,Zhang:2018:DDL:3159652.3159688}, which generate item hash codes differently depending on whether they are cold-start items or not. 
NeuHash-CF is formulated as a variational autoencoder architecture, where both user and item hash codes are sampled from learned Bernoulli distributions to enforce end-to-end trainability. We presented a comprehensive experimental evaluation of NeuHash-CF in both standard and cold-start settings, where NeuHash-CF outperformed state-of-the-art approaches by up to 12\% NDCG and 13\% MRR in cold-start recommendation (up to 4\% in both NDCG and MRR in standard recommendation settings). In fact, the ranking performance of NeuHash-CF on 16 bit hash codes is better than that of 32-64 bit state-of-the-art hash codes, thus resulting in both a significant effectiveness increase, but also in a 2-4x storage reduction. Analysis of our results showed that the largest performance difference between hashing-based and real-valued approaches occurs for users interested in the least popular items, and for the group of users with the highest number of rated items. 
Future work includes extending the architecture to accept richer item and user representations, such as \cite{Hansen0ASL19,WangZLLZS20,RashedGS19,CostaD19}.

\clearpage

\bibliographystyle{ACM-Reference-Format}
\balance
\bibliography{main.bib}

\end{document}